\documentclass{aastex62}

\received{May 20, 2020}
\submitjournal{ApJ}

\shorttitle{Black Hole Spin Mass-Energy Parameters and the Spin Energy Reservoir}
\shortauthors{Daly} 
\usepackage{longtable}
\begin{document}

\title{Black Hole Spin Energy Contribution to Black Hole Mass and the Spin 
Energy Reservoir}

\correspondingauthor{Ruth Daly}
\email{rdaly@psu.edu}

\author[0000-0000-0000-0000]{Ruth A. Daly}

\affil{Penn State University, Berks Campus, Reading, PA 19608, USA}
\affil{Center for Computational Astrophysics, Flatiron Institute, 162 Fifth Avenue, New York, NY 10010, USA}

\nocollaboration

\begin{abstract}
The mass of a black hole is the sum of the irreducible mass 
and the mass associated with the rotational or spin energy of the black hole. 
The contribution of spin energy (divided by $c^2$) to the total black hole mass 
is studied here for four samples of sources including 
576 LINERs, 100 classical double radio sources, 80 relatively local AGN, and 
102 measurements of four stellar mass X-ray binary systems. 
The spin mass-energy of a black hole may be extracted causing the mass of the 
black hole to decrease. The 
ratio of spin mass-energy to black hole mass ranges from about 
ten to thirty percent for the sources studied here, where the maximum possible 
value of this quantity is close to thirty percent. Typical fractions of the black hole
mass available for extraction for the samples studied are about $0.2 \pm 0.1$.  
The spin energy of black holes represents a major reservoir 
that when tapped may impact the immediate and 
extended black hole environment, decrease the black hole mass, and perhaps modify 
relationships between black hole mass and galaxy properties. These results are 
consistent with expectations based on other observations and numerical simulations. 
\end{abstract}

\keywords{black hole physics -- galaxies: active}

\section{Introduction} \label{sec:intro}
Black holes are ubiquitous in the universe. Supermassive black holes reside at the 
centers of galaxies and stellar-mass black holes populate galaxies. The two primary 
characteristics that describe an astrophysical black hole are the mass and spin 
of the hole. Typically in astrophysical contexts, the mass of a black hole 
is determined by the dynamics 
and properties of matter and light in the vicinity of the black hole. 
The black hole mass that will be measured is the total or gravitational 
black hole mass, $M$, which is the 
sum of the irreducible mass, $M_{irr}$, and the mass-energy associated with the 
spin of the hole, $M_{spin}$: $M = M_{irr} + M_{spin}$, where 
$M_{spin} = E_{spin} c^{-2}$, $E_{spin}$ is the spin energy of the black hole, 
and $c$ is the speed of light
(e.g. Christodoulou 1970; Bardeen, Press, \& Teukolsky 1972; Thorne et al. 1986; 
Berti, Cardoso, \& Starinets 2009). 
The mass, $M$, is also referred to as the dynamical mass, $M_{dyn}$, since it  
is the total black hole mass that will be inferred by dynamical studies. 

The irreducible mass of an isolated black hole can not be reduced or decreased, 
but part or all of the mass-energy associated with black hole spin can be 
extracted (Penrose 1969), thereby decreasing the total mass of the hole 
(e.g. Rees 1984; Blandford 1990). Collimated outflows from 
supermassive black holes associated with 
active galactic nuclei (AGN) and stellar-mass black holes associated with 
X-ray binaries are likely to be powered, at least in part, by black hole spin 
(Blandford \& Znajek 1977; MacDonald \& Thorne 1982; Phinney 1983; 
Begelman, Blandford, \& Rees 1984; Blandford 1990; Daly 1994; Moderski, Sikora, 
\& Lasota 1998; Meier 1999; Koide et al. 2000; 
Punsly 2001; De Villiers, Hawley, \& Krolik 2003; Gammie et al 2004; 
Komissarov \& McKinney 2007; Beckwith, Hawley, \& Krolik 2008; 
King, Pringle, \& Hofmann 2008; Miller et al. 2009; Tchekhovskoy et al. 2010; 
Gnedin et al. 2012; King et al. 2013; Ghisellini et al. 2014; Yuan 
\& Narayan 2014; Gardner \& Done 2018; Krause et al. 2019; 
Reynolds 2019). In this case, the spin energy extracted during the outflow will 
cause the black hole mass to decrease. A source that undergoes multiple outflow 
events could significantly drain the spin energy of the hole and thereby 
decrease the black hole mass. 
The amount of spin energy extracted during outflow events have been estimated 
for radio sources with large-scale outflows such as 
FRI sources in galaxy-cluster environments 
(McNamara et al. 2009; Daly 2009a,b, 2011) and FRII sources 
(Daly 2009a,b; 2011); FRI sources are extended radio sources that are 
"edge-darkened" while FRII sources, also known as classical doubles, 
are "edge-brightened" (Fanaroff \& Riley 1974). 
In addition, the fraction of the spin energy extracted
per outflow event has been estimated for FRII sources (Daly 2011; Daly 2020), and 
is roughly a few percent. 
Thus, in models in which collimated outflows from the vicinity of
a black hole are powered by black hole spin, the spin and spin-energy of the hole
are expected to decrease as a result of the outflow. 

The fact that the mass-energy associated with black hole spin may 
be extracted, modified, or reduced, 
and thus that the total or dynamical mass of a black hole can be 
reduced may introduce 
dispersion in relationships between black hole mass and properties of the host galaxy 
(e.g. Kormendy 
\& Richstone 1995; Ferrarese \& Ford 2005; Kormendy \& Ho 2013; Shankar 2013; 
Sesana et al. 2014; Zubovas \& King 2019; King \& Nealon 2019; King \& Pounds 2015). 
If black hole spin evolves with 
redshift, this is likely to  cause an evolution in these relationships and 
their dispersion. Additionally, black hole spin is expected to evolve with redshift 
as a result of the merger and accretion history of the black hole 
(e.g. Hughes \& Blandford 2003; Gammie et al. 2004; 
Volonteri et al. 2005, 2007; King \& Pringle 2006, 2007; King et al. 2008;
Berti \& Volonteri 2008; Ghisellini et al. 2013). 
Thus, the study of black hole spin evolution provides insight into the 
merger and accretion history of supermassive black holes. 
Black hole spin may depend upon galaxy type or environment (e.g. Sesana et al. 2014; 
Antonini et al. 2015; King \& Pounds 2015; Barausse et al.  2017; 
King \& Nealon 2019), which may lead to environmental changes in the relationship between 
black hole mass and galaxy properties, 
or a change in the dispersion of relationships (e.g. Zubovas \& King 2012). 
The dispersion introduced 
may be complex and will depend upon the initial spin and irreducible 
mass of the black hole, the processes responsible for spinning up the hole 
such as accretion 
or mergers, processes which tap or reduce the spin of the hole, 
and the complex interaction of feedback, accretion, outflows, and other processes 
associated with the black hole, which are likely to play a role in determining the spin 
and thus spin mass-energy and dynamical mass of the hole 
(e.g. Belsole et al. 2007; Worrall 2009; Voit et al. 2015; 
Hardcastle \& Croston 2020). In addition, 
it is likely that some sources undergo multiple outflow events (e.g. Hardcastle et al. 
2019; Bruni et al. 2019, 2020; Shabala et al. 2020). 

The distinction between dynamical mass, spin mass-energy, and irreducible mass of a  
black hole is also important when comparing empirically determined quantities with 
theoretically predicted quantities, such as those indicated by numerical simulations. 
Numerical simulations predict the expected black hole spin and mass evolution in 
the context of different black hole merger and accretion histories (e.g. 
King et al. 2008; Volonteri et al. 2013; 
Dubois, Volonteri, \& Silk 2014; Sesana et al. 2014; Kulier et al. 2015). 
A comparison of simulation results with empirically determined results  
provides an important diagnostic of the merger and accretion histories of 
black holes located at the centers of galaxies. 

The number of available black hole spin values, and therefore black hole spin
energies, has recently increased substantially. 
The development of the "outflow method" of empirically determining black hole 
spin and accretion disk properties developed and described by Daly (2016, 2019) 
(hereafter D16 and D19) and 
Daly et al. (2018), allow the empirical determination of the black hole spin 
function, spin, and, accretion disk properties such as the 
mass accretion rate and disk magnetic field strength for hundreds of sources.  
D19 showed that the fundamental equation that describes an outflow 
powered at least in part by black hole spin,
$L_j \propto B_p^2 M_{dyn}^2 f(j)$ (e.g. Blandford \& Znajek 1977; 
Meier 1999; Tchekhovskoy et al. 2010; Yuan \& Narayan 2014) is 
separable and may be written as 
\begin{equation}
(L_j/L_{Edd}) = g_j ~(B/B_{Edd})^2 ~(f(j)/f_{max})
\end{equation}
(see eq. 6 from D19); here $B_p$ is the poloidal component of the accretion disk magnetic field, 
$B$ is the magnitude of disk magnetic field, $B_{Edd}$ is the Eddington magnetic 
field strength (e.g. Rees 1984; Blandford 1990; Dermer et al. 2008; D19), 
$B_{Edd} \approx 6 \times 10^4 (M_{dyn}/10^8 M_{\odot})^{-1/2}$ G, $f(j)$ 
is the spin function (discussed in more detail in section 3.1), $f_{max}$ 
is the maximum value of this function, and 
$g_j$ is the normalization factor for the beam power $L_j$ in units of the 
Eddington Luminosity, $L_{Edd}$, $(L_j/L_{Edd})(max) = g_j$. 
Note that the ratio $(B_p/B)^2$ is absorbed into the 
normalization factor $g_j$, thus $g_j$ may depend upon AGN type, as 
discussed in sections 3.1 and 4 of D19. 

The black hole spin values obtained by D19 are similar to those obtained 
with independent methods such as those summarized by Reynolds (2019);  
see also, for example, Gnedin et al. (2012), Patrick et al. (2012),   
King et al. (2013), Walton et al. (2013), Wang et al. (2014), 
Garc\'ia et al. (2015), Mikhailov et al. 
(2015, 2019), Vasudevan et al. (2016), Piotrovich et al. (2017, 2020), and 
Mikhailov \& Gnedin (2018). Very good agreement was found for 
spin determinations obtained independently using the outflow method and 
the X-ray reflection method (e.g. Fabian et al. 1989; 
Iwasawa et al. 1997; Miller et al. 2002; Reynolds 2019) 
for six AGN and one GBH for which a comparison of values was possible. 
The high spin values obtained are also consistent
with expectations based on AGN luminosities 
(e.g. Sun \& Malkan 1989; Davis \& Laor 2011; Wu et al. 2013; Trakhtenbrot 2014; 
Brandt \& Alexander 2015; Trakhtenbrot, Volonteri, \& Natarajan 2017). 
So, the expectation is that some significant fraction of black holes are likely 
to have high spin; there may also be a population of black holes with lower 
spin, and, of course, the spin is likely to be an evolving quantity. 

Here, four samples of sources that allow the spin mass-energy, irreducible black 
hole mass, and total black hole mass to be studied are considered. 
The samples are described in section 2. The method of obtaining spin mass-energy 
relative to the irreducible or dynamical mass is described in section 3. 
The results are described in section 4, discussed in section 5, and summarized 
in section 6. 

\section{Data} \label{sec:data}

Four samples are considered here. The samples include 576 LINERs from Nesbit \& Best 
(2016) (hereafter NB16), 80 AGN that are compact radio sources from 
Merloni et al. (2003) (hereafter M03), 100 FRII from Daly (2016, 2019), 
(hereafter D16 and D19, respectively) 
and 102 observations of four Galactic Black Hole systems (GBH), 
which are stellar mass X-ray binaries, and three of the four GBH have multiple 
simultaneous radio and X-ray observations (Merloni et al. 2003; 
Gallo et al. 2006; Corbel et al. 2008, 2013; Saikia et al. 2015, hereafter S15). 
These four samples were studied by D19 who reported spin 
functions, spin values, and accretion disk magnetic field strengths in 
dimensionless and physical units for each source and observation. 
The spin functions and spin values reported by D19 are used here to 
study the contribution of spin energy to the total black hole mass of each system, 
as discussed in section 3.1. The results are summarized in Table 1. 

\begin{figure}
    \centering
    \includegraphics[width=100mm]{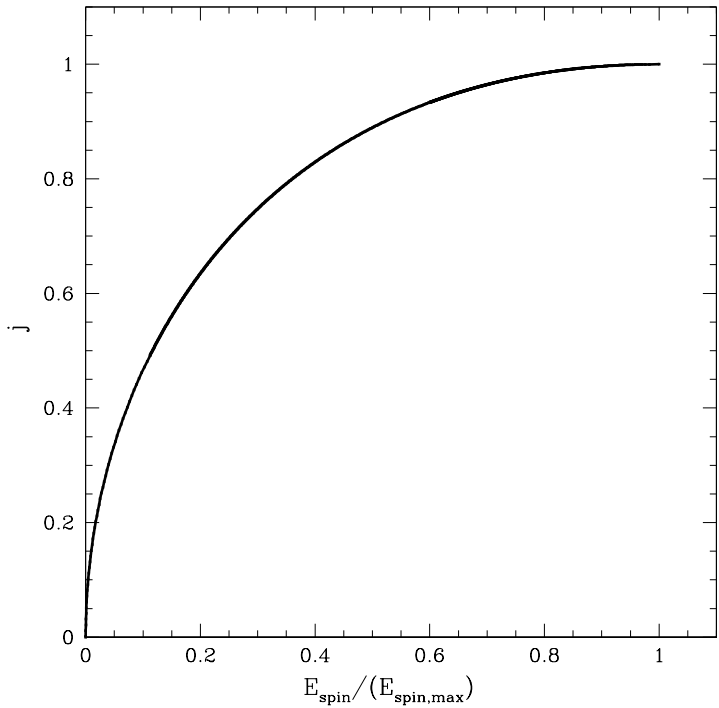}
\caption{The black hole spin $j$ is shown versus the black hole spin energy in 
units of the maximum possible value of the spin energy, $E_{spin,max}.$}
		  \label{fig:FEspinoverEspinmax}
    \end{figure} 

\clearpage   
\section{Method} \label{sec:method}

\subsection{Spin Mass-Energy Contribution to Black Hole Mass}
\label{3.1}

Relationships between the dimensionless black hole spin, $j$, 
black hole mass, $M$, and the mass-energy associated with black hole 
spin, $M_{spin}$, are   
discussed, for example, by  Misner, Thorne, \& Wheeler (1973) and  
Thorne et al. (1986). The black hole spin $j$ is defined in the usual way, 
$j \equiv Jc/(G M^2)$, where $J$ is the spin angular momentum of the hole  
and $G$ is Newton's constant; in other work, $j$ is sometimes represented with the 
symbol $a$ or $a_*$. From this work, we obtain the following set of equations:  
\begin{equation}
  M \equiv M_{dyn} = M_{irr} + E_{spin} c^{-2} = M_{irr} + M_{spin}  
\end{equation}
and 
\begin{equation}
    M_{irr} = M_{dyn} \left({1 + (1-j^2)^{1/2} \over 2}\right)^{1/2}
\end{equation}
so
\begin{equation}
{M_{spin} \over M_{dyn}} = 1 - \left({M_{irr} \over M_{dyn}}\right)
\end{equation}
and 
\begin{equation}
{M_{spin} \over M_{irr}} = \left({M_{dyn} \over M_{irr}}\right) -1~.
\end{equation}

Empirically, the black hole spin $j$ for each source is obtained from the 
spin function $f(j)/f_{max}$, where $F \equiv \sqrt{f(j)/f_{max}} = j (1+\sqrt{1-j^2})^{-1}$, following the procedure described by D19. 

\begin{figure}
    \centering
    \includegraphics[width=100mm]{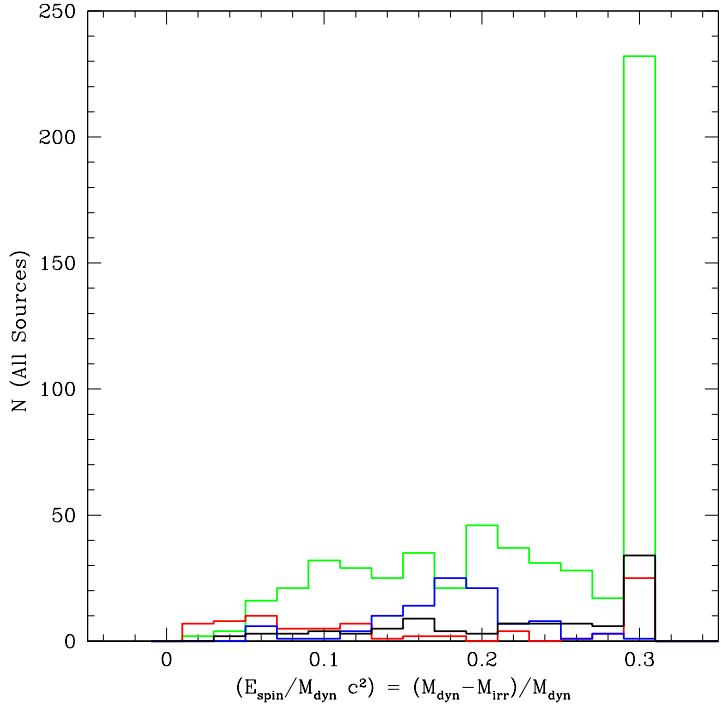}
\caption{Histograms of $M_{spin}/M_{dyn}$ are
shown for each of the four samples studied. Here and throughout the 
paper, the 576 LINERs from NB16 are shown in green; the 100 FRII sources 
from D16 and D19 are shown in black; the 80 AGN from M03 are shown in red; 
and the 102 GBH from S15 are shown in blue. Sources with j = 0.90, 0.95, 
and 1 get mapped to values of $M_{spin}/M_{dyn}$ of 0.15, 0.19, 
and 0.29, respectively. As discussed by D19 and in section 4, the conversion from 
the empirically determined spin function to black hole spin requires that 
sources with spin functions greater than unity get mapped to spin values of 1, 
causing a cut-off for values of $M_{spin}/M_{dyn}$ greater than about 0.29.}
		  \label{fig:ANF1}
    \end{figure} 
    
\begin{figure}
    \centering
    \includegraphics[width=100mm]{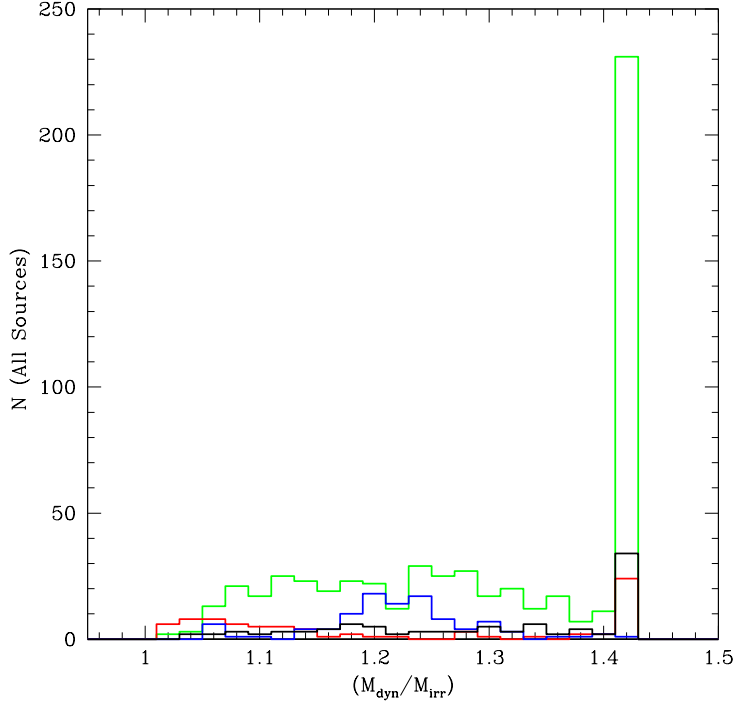}
\caption{Histograms of $M_{dyn}/M_{irr}$ are
shown for each of the four samples studied here with the colors as in 
Fig. \ref{fig:ANF1}. Sources with j = 0.90, 0.95, and 1 get 
mapped to values of $M_{dyn}/M_{irr}$ of 1.18, 1.23, and 1.41, respectively. 
As discussed in section 4, 
the conversion from the empirically determined spin function to black hole 
spin requires that sources with spin functions greater than unity get mapped 
to spin values of 1, causing a cut-off for values of $M_{dyn}/M_{irr}$ greater 
than about 1.41.}
		  \label{fig:ANF3}
    \end{figure} 

The relationship between black hole spin and spin mass-energy is illustrated in 
Fig. 1. The maximum value of $M_{spin}$ can be obtained 
from eqs. (3) and (4) with $j=1$, which indicate 
$(M_{spin}/M_{dyn})(max) \simeq 0.29$ and 
$(M_{spin}/M_{irr})(max) \simeq 0.41$. Normalizing the spin mass-energy with these 
maximum values, Fig. 1 illustrates the relationship between the 
normalized spin energy and black hole spin $j$. The spin energy $E_{1/2}$
is about half of the maximum value for a spin of about 
0.9; the spin energy $E_{1/4}$ is about one quarter of the maximum value 
when the spin is about 0.7; 
and the spin energy $E_{1/10}$ is about one tenth of the maximum value 
when the spin is about 0.5.
It is clear that the relationship between spin energy and spin is quite 
non-linear, and 
relatively high spin values of 0.5 and 0.7 indicate relatively low spin energy values 
of only 1/10 and 1/4, respectively, of the maximum possible spin energy. 
Another way to 
state this is that relatively low values of spin energy indicate substantial values of 
black hole spin $j$. Thus, if a black hole has any spin energy at all, it is 
expected to have a value of $j$ substantially different from zero. 

\subsection{Parameter Uncertainties}
\label{3.2}

For the sources that will be studied here, the dynamical mass, $M_{dyn}$,  
and the black hole spin function, $F^2$, or spin $j$, and their uncertainties  
are presented by D19. The quantities and their uncertainties 
to be determined here are the ratios  $(M_{dyn}/M_{irr})$ (or it's inverse), 
$(M_{spin}/M_{dyn})$, $(M_{spin}/M_{irr})$, and $M_{spin}$. 
To compute the uncertainty of each quantity, it is noted that 
$F = 0.5~ j ~(M_{dyn}/M_{irr})^2$.

Using eq. (3), it is easy to show that $\delta \rm{Log}(M_{spin}/M_{dyn}) = \delta \rm{Log}(M_{irr}/M_{dyn})$. Eq. (4) indicates that 
$\delta \rm{Log}(M_{spin}/M_{irr}) = \delta \rm{Log}(M_{dyn}/M_{irr})$, so 
$\delta \rm{Log}(M_{spin}/M_{irr}) = \delta \rm{Log}(M_{irr}/M_{dyn})$. 
Given that $(M_{irr}/M_{dyn})^2 = 0.5~ (j/F)$, it can be shown that  
$\delta \rm{Log}(M_{irr}/M_{dyn}) \leq (2)^{-1/2} \delta \rm{Log}(F)$, since 
$(\delta j/j) \leq (\delta F/F)$, which follows from 
$(\delta j/j) = \sqrt{(1-j^2)} (\delta F/F)$. Here, 
$\delta \rm{Log}(M_{irr}/M_{dyn}) \simeq 0.7 ~\delta \rm{Log}(F)$ is used.
Eq. (2) indicates that $\delta \rm{Log}(M_{irr}) = [(\delta \rm{Log}(M_{dyn}))^2
+ 0.25 ~(j F)^2 (\delta \rm{Log}(F))^2]^{1/2}$. The second term is small compared with 
the first term, and thus is 
negligible, so this becomes $\delta \rm{Log}(M_{irr}) \simeq \delta \rm{Log}(M_{dyn})$. 
This follows because the dynamical mass is simply related 
to the Eddington luminosity, $L_{Edd} \simeq 1.3 \times 10^{38}(M_{dyn}/M_{\odot}) 
\hbox{ erg s}^{-1}$, or $M_{dyn} \propto L_{Edd}$, and the uncertainty of 
$\delta \rm{Log}(F)$ is typically about half of the uncertainty of $\delta \rm{Log}(L_{Edd})$
for each sample (see section 3.3 of D19). It is easy to see that 
$\delta \rm{Log}(M_{spin}) = [(\delta \rm{Log}(M_{dyn}))^2 + (\delta \rm{Log}(M_{spin}/M_{dyn})^2)]^{1/2}$; combining the values of 
$\delta \rm{Log}(M_{spin}/M_{dyn})$ given above and the values of 
$\delta \rm{Log}(M_{dyn})$, which is equal to 
$\delta \rm{Log}(L_{Edd})$, listed by D19,  
$\delta \rm{Log}(M_{spin})$ can be obtained.

Applying these expressions, the uncertainty per source is estimated to be: 
$\delta \rm{Log}(M_{irr}/M_{dyn}) 
= \delta \rm{Log}(M_{spin}/M_{dyn}) = \delta \rm{Log}(M_{spin}/M_{irr}) \simeq$ 
0.11, 0.13, 0.11, and 0.05 for the 576 LINERs, 80 M03 AGN, 100 D19 FRII AGN, 
and the 102 GBH measurements, respectively. If any of these ratios are  
represented as x,
(e.g. $x = M_{irr}/M_{dyn}$, or $x = M_{spin}/M_{dyn}$, or $x = M_{spin}/M_{irr}$), 
then $\delta x \simeq 0.25 ~x$ for 
$\delta \rm{Log}(x) \simeq 0.11$, since $(\delta x/x) = \rm{Ln}(10)~\delta \rm{Log}(x)$. 
Using the expression provided above, the uncertainty per source is estimated to be 
$\delta \rm{Log}(M_{spin}) \simeq$ 
0.37, 0.52, 0.37, and 0.05 for the 576 LINERs, 80 M03 AGN, 100 D19 FRII AGN, 
and the 102 GBH measurements, respectively. Here, the uncertainty of the 
GBH mass term is neglected since it is typically 
small compared with $\delta \rm{Log}(M_{spin}/M_{dyn})$, as discussed by D19. 
    
\section{Results} 
\label{sec:results} 

Equations (2) - (5) are applied to empirically determine the mass-spin energy, $M_{spin}$,  
of each black hole and related quantities. These are summarized Table 1 for each of the 
four samples, and for each of the GBH, which have multiple observations per source. 
Histograms showing results of the ratio of black hole spin mass-energy to 
dynamical mass, and dynamical mass relative to irreducible mass are 
shown in Figs. \ref{fig:ANF1} and \ref{fig:ANF3}. To get some perspective on 
the contributions to the histograms from sources at different redshift, quantities 
are shown versus redshift is Figs. \ref{fig:NF4} to \ref{fig:F7}. 

As indicated by Fig. \ref{fig:ANF1}, all four samples include sources with a broad 
range of values of $M_{spin}/M_{dyn}$, and the three AGN samples exhibit a 
peak at the maximum possible value, about 0.29, of this parameter. 
One factor that contributes to this peak 
is that the translation of the spin function, $F^2$, to spin, $j$, requires 
that sources with a spin function larger than one be interpreted as having a 
spin of one (e.g. D19). 
To consider the properties of $M_{spin}/M_{dyn}$ 
in more detail, each of the four samples were split 
into those sources with a maximum value of $M_{spin}/M_{dyn}$, that is, those with 
$M_{spin}/M_{dyn} \geq 0.29$, and the rest of the sample. 
For the 80 M03 sources, 100 D19 FRII sources, 576 NB16 LINERs, and 102 S15 GBH, 
these maximum 
mass-spin energy sources accounted for about 32 \%, 35 \%, 40 \%, and 2 \%
of the sample sources, respectively. The remaining 54 M03 sources, 
65 D19 FRII sources, 345 LINERs, and 100 GBH have
mean values of $M_{spin}/M_{dyn}$ of about 
$0.10 \pm 0.07$, $0.18 \pm 0.07$, $0.17 \pm 0.06$, 
and $0.17 \pm 0.04$, respectively. For the AGN, the redshift distributions of 
each sample are quite different as will be discussed below, 
but each of the samples includes sources 
with a broad range of values of $M_{spin}/M_{dyn}$, and sources with maximum values of 
this quantity. These results are mirrored in the distributions of $M_{dyn}/M_{irr}$, the ratio of the
total or dynamical black hole mass to the irreducible mass, which are shown in Fig. 
\ref{fig:ANF3}. As indicated by eq. (2), the maximum possible value of this
quantity is $\sqrt(2) \simeq 1.41$. 

As discussed in section 1, uncertainties in the ratio of $(B_p/B)$ lead to 
uncertainties in $g_j$, which will impact the spin functions and spins, since 
$F \propto g_j^{-1/2}$. This may account for sources with 
large values of $F$, as discussed by D19. Other effects, such as fluctuations 
introduced by non-simultaneous radio and X-ray observations, which would 
only impact the AGN, could also produce
values of $F$ that are greater than one (D19). To conservatively estimate the 
reservoir of mass-energy associated with black hole spin, the impact of removing 
sources with values of $\rm{Log}(F)$ that exceed zero at more than one sigma   
is considered here. This leaves 10 M03 sources, 24 FRII sources, 
121 LINERs, and two S15 GBH in the bin with the 
highest possible value of $M_{spin}/M_{dyn}$ and $M_{dyn}/M_{irr}$, or about 
16 \%, 27 \%, and 26 \% of the remaining M03, FRII, and LINERs sources, respectively. 
Thus, after removing sources with values of $\rm{Log}(F) > 0$ at more 
than one sigma, the strong peak of maximally spinning holes persists for 
FRII AGN and LINERs, and the distributions include a tail of sources that 
are roughly evenly distributed across $M_{spin}/M_{dyn}$ and $M_{dyn}/M_{irr}$.  
For the M03 sample of local AGN, there is no longer a prominent peak of maximally 
spinning holes, and $M_{spin}/M_{dyn}$ and $M_{dyn}/M_{irr}$ are roughly evenly distributed 
between the minimum and maximum values. This suggests that there may be two populations 
of supermassive black holes: a population of maximally spinning holes and a population 
of holes with a roughly even distribution of spin energy relative to dynamical black 
hole mass. Potential selection effects relevant to these results are discussed in section 5. 
\clearpage

\begin{table*}
\begin{minipage}{165mm}
\caption{Mean Value and Standard Deviation of Black Hole Spin Mass-Energy Parameters
{%
\footnote{Values in parentheses indicate the estimated uncertainty per source, as discussed in section 3.2.}}}   
\label{tab:mean}        
\begin{tabular}{llllllllll}   
\hline\hline                    
(1)&(2)&(3)&(4)&(5)&(6)&&(7)&(8)&\\
	&		&	&&&$\rm{Log}$&&$	\rm{Log}$&$	\rm{Log}$	\\
	
		&		&	N{%
\footnote{N is the number of sources in the sample for AGN, and the number of simultaneous radio and X-ray measurements for GBH.}}	&$M_{dyn}/M_{irr}$&$	{M_{spin}/M_{irr}}$&$({M_{spin}/M_{dyn}})$&&$(M_{spin})$&$(M_{dyn})$	\\
&&&&&&&$(M_{\odot})$&$(M_{\odot})$\\
\hline
NB16	&	AGN	&	576	&$	1.30	\pm	0.12	$&$	0.30	\pm	0.12	$&$	-0.69	\pm	0.20(0.11)	$&&$	7.38	\pm	0.52(0.37)	$&$	8.07	\pm	0.50(0.35)	$	\\
D19	&	AGN	&	100	&$	1.29	\pm	0.12	$&$	0.29	\pm	0.12	$&$	-0.70	\pm	0.21(0.11)	$&&$	8.41	\pm	0.50(0.37)	$&$	9.11	\pm	0.41(0.35)	$	\\
M03	&	AGN	&	80	&$	1.21	\pm	0.16	$&$	0.21	\pm	0.16	$&$	-0.93	\pm	0.38(0.13)	$&&$	6.91	\pm	0.89(0.52)	$&$	7.84	\pm	0.81(0.50)	$	\\
S15	&	GBH	&	102	&$	1.22	\pm	0.07	$&$	0.22	\pm	0.07	$&$	-0.77	\pm	0.14(0.05)	$&&$	0.06	\pm	0.18(0.05)	$&$	0.83	\pm	0.09	$	\\
\hline 
\hline 
GX339-4	&	GBH{%
\footnote{Results for individual GBH are included in the second half of the table; for A06200 the dispersion of each quantity is taken to be that of the full GBH sample. Mass uncertainties for GBH are discussed by D19.}}  
&	76	&$	1.21	\pm	0.05	$&$	0.21	\pm	0.05	$&$	-0.77	\pm	0.10	$&&$	0.01	\pm	0.10	$&$	0.78	$	\\
V404Cyg	&	GBH	&	20	&$	1.28	\pm	0.06	$&$	0.28	\pm	0.06	$&$	-0.67	\pm	0.07	$&&$	0.33	\pm	0.07	$&$	1.00	$	\\
J1118+480	&	GBH	&	5	&$	1.07	\pm	0.01	$&$	0.07	\pm	0.01	$&$	-1.20	\pm	0.02	$&&$	-0.32	\pm	0.02	$&$	0.88	$	\\
A06200	&	GBH	&	1	&$	1.29	\pm	0.07	$&$	0.29	\pm	0.07	$&$	-0.65	\pm	0.14	$&&$	0.17	\pm	0.18	$&$	0.82	$	\\
\hline 
\end{tabular}
\end{minipage}
\end{table*}

\section{Discussion} 
\label{sec:discussion}
The outflow method of determining black hole spin and accretion disk 
properties proposed and 
applied by D16 and D19 allows empirical estimates of the spin mass-energy, 
$M_{spin}$, and $M_{spin}$ relative to the dynamical mass, 
$M_{dyn}$, or the irreducible mass, $M_{irr}$, of the black hole. 
In addition to considering histograms of quantities, it is helpful to view the 
redshift distributions of the sources to get a sense of contributions to the 
histograms from sources at different redshift. Since GBH are located in 
the Galaxy, these are not included in the redshift distributions. The M03 AGN 
are quite local, and include sources with a wide range of intrinsic 
parameters. The NB16 LINERs extend to a redshift of about 0.3, with this cutoff 
imposed by NB16. The FRII sources have redshifts between about zero and 2, 
and are selected from the 178 MHz radio flux limited
3CRR sample, described by Laing, Riley, \& Longair (1983). 
It is easy to see the impact of missing lower luminosity sources 
as redshift increases. Note that the upper envelope of the distributions 
provides a guide as to how parameters that described sources with the largest beam 
power evolve with redshift. Beam powers for FRII sources are obtained using the 
strong shock method (e.g. Daly 1994; O'Dea et al. 2009); the method is 
discussed in detail by D19. For the LINERs, M03, and 
GBH samples, empirically determined beam powers are obtained using eq. (4) of 
Daly et al. (2018) or eq. (4) of D19 using best fit parameters obtained by mapping the 
fundamental plane to the fundamental line of black hole activity (Daly et al. 2018; D19). 
For the FRII sources, the redshift distribution of quantities discussed here 
as well as those of several additional parameters such as the fraction of spin 
energy and angular momentum extracted per outflow event are discussed in detail by 
Daly (2020). 

\begin{figure}
    \centering
    \includegraphics[width=100mm]{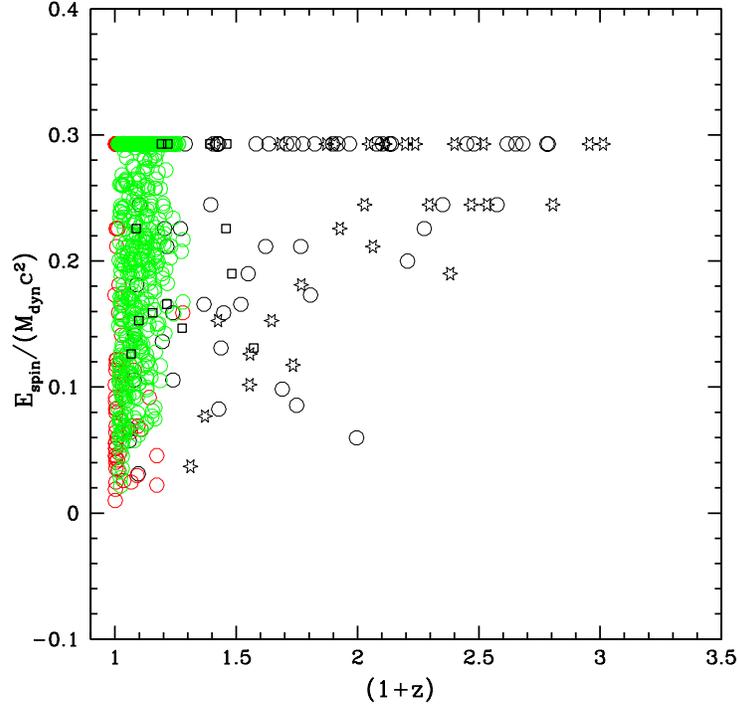}
\caption{Values of $M_{spin}/M_{dyn}$ are shown vs. $(1+z)$ for the three AGN 
samples with the same color coding as in Fig. \ref{fig:ANF1}, with 80 AGN 
from M03 shown in red, 576 LINERs from NB16 shown in green, and 100 FRII AGN from 
D16 and D19 shown in black; for the FRII sources black squares indicate LEG, 
circles indicate HEG, and stars indicate Q. This color scheme and symbols apply 
to all subsequent figures. A plot of $M_{spin}/M_{irr}$ vs. $(1+z)$ is very 
similar except that $M_{spin}/M_{irr}$ extends to about 0.41. }
		  \label{fig:NF4}
    \end{figure}

\begin{figure}
    \centering
    \includegraphics[width=100mm]{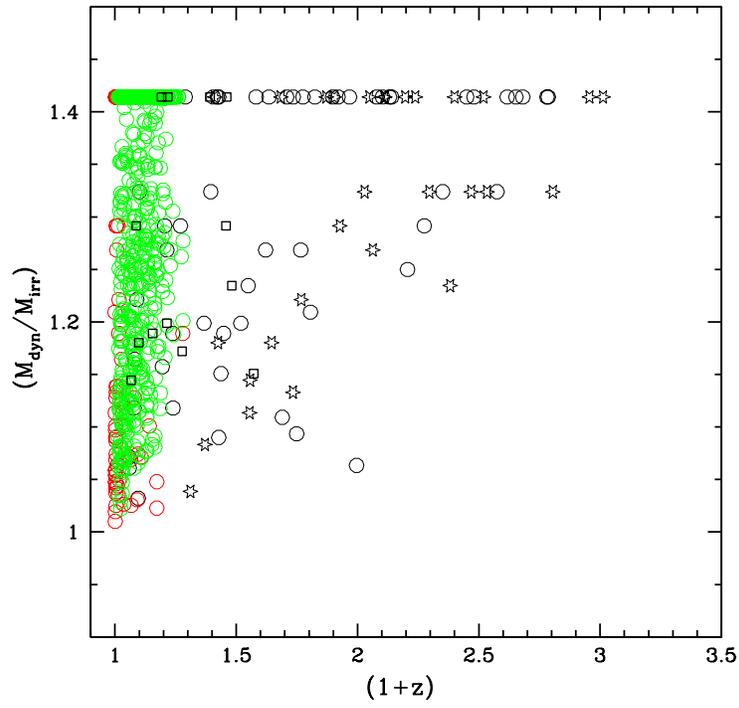}
\caption{Values of $M_{dyn}/M_{irr}$ are shown vs. $(1+z)$ for the three AGN 
samples with the same colors and symbols as in Fig. \ref{fig:NF4}. These results are discussed in section 5. }
		  \label{fig:NF5}
    \end{figure}

The ratio $M_{spin}/M_{dyn}$ and $M_{dyn}/M_{irr}$ are shown as functions of 
redshift in Figs. \ref{fig:NF4} and \ref{fig:NF5}. For the M03 and NB16 samples, 
all of the sources are at relatively low redshift, including those with maximum 
values of the spin mass-energy parameters. 
The FRII sources have a broad range of parameters at 
all redshift, with the impact of the flux limited nature of the 3CRR survey 
likely accounting for the dearth of low spin mass-energy sources at high redshift. 
This can be addressed empirically with future studies that include lower luminosity 
sources at high redshift. In terms of the results presented here, the lack of 
sources with lower values of $M_{spin}/M_{dyn}$ and $M_{dyn}/M_{irr}$ could be 
due to a selection effect whereby these sources are not included in the samples 
studied. 

\begin{figure}
    \centering
    \includegraphics[width=100mm]{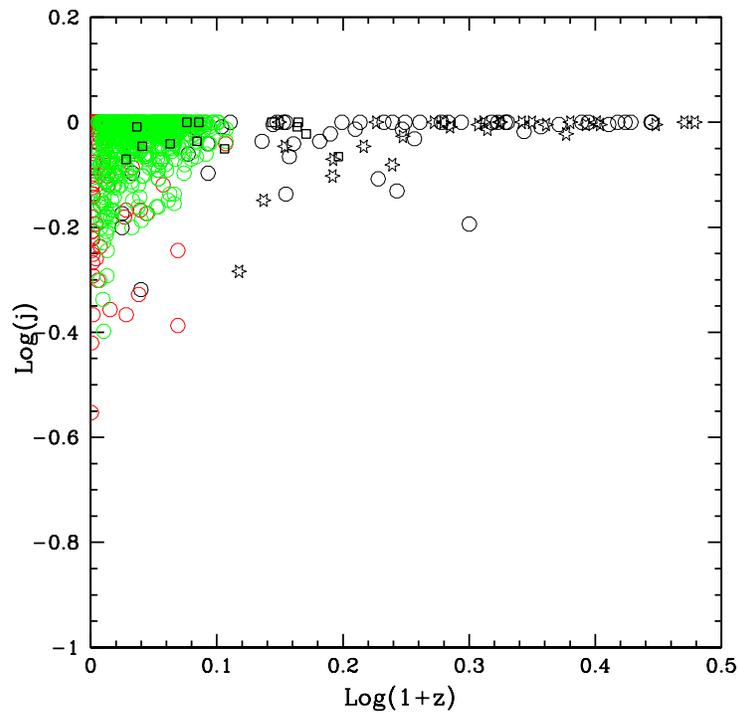}
\caption{Values of $\rm{Log}(j)$ are shown vs. $(1+z)$ for the three AGN samples with 
the same colors and symbols as in Fig. \ref{fig:NF4}.}
		  \label{fig:F8}
    \end{figure}

\begin{figure}
    \centering
    \includegraphics[width=100mm]{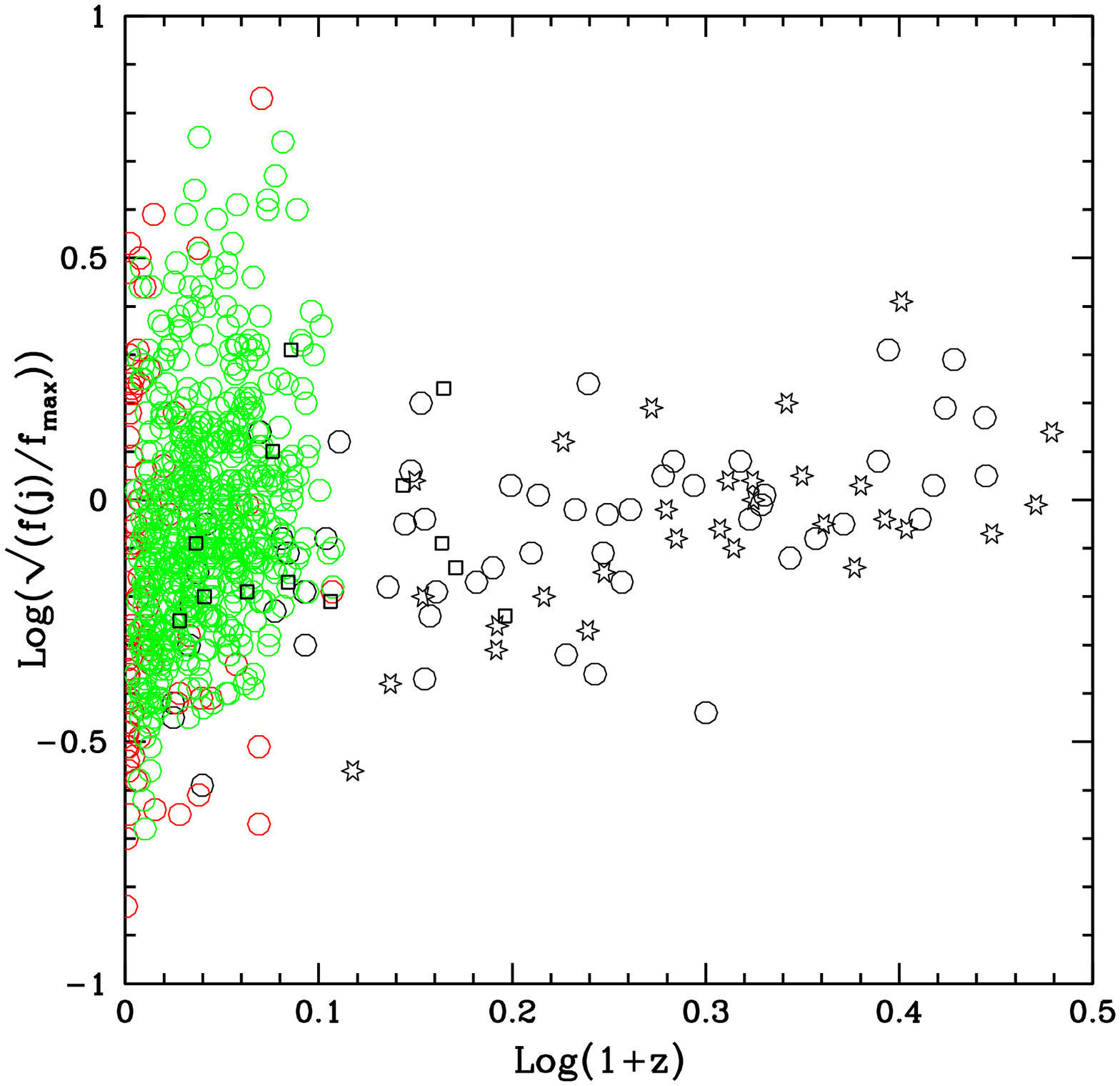}
\caption{Values of $\rm{Log}(F)$ are shown vs. $\rm{Log}(1+z)$ for the three AGN samples with 
the same colors and symbols as in Fig. \ref{fig:NF4}. Assuming zero uncertainty 
for $g_j$, the uncertainty of $\rm{Log}(F)$ is about 0.15 for the LINERs and 
FRII sources, and 0.19 for the M03 local sample of AGN (D19).}
		  \label{fig:F7}
    \end{figure}

Fig. \ref{fig:F8} illustrates how a rather 
small range of values of black hole spin $j$ leads to a broad range of 
values of $M_{spin}/M_{dyn}$ and $M_{dyn}/M_{irr}$, as discussed in section 3.1.
The values of spin $j$ are obtained from values of the spin function $F$, 
shown in Fig. \ref{fig:F7}. As discussed in detail by D19, the conversion of 
$F$ to $j$ requires that values of $F \geq 1$ get mapped in values of $j=1$. 
The impact of excluding sources with values of $\rm{Log}(F)$ that exceed zero 
by more than one sigma is discussed in section 4. As discussed in that section, 
variations in the value of $g_j$, which includes the term $(B_p/B)^2$, may well 
depend upon AGN type and could artificially increase the value of the spin function.
Even after accounting for this effect, there remains a significant population 
of maximally spinning black holes for the LINERs and FRII sources studied, 
as well as black holes with lower spin mass-energy parameter values. 

The FRII AGN have redshifts between about zero and two, and 
exhibit a broad range of spin mass-energy parameters including 
a significant number of sources with maximal spin mass-energy, as discussed 
in section 4. The envelope of the distribution of these sources is comprised of 
the most powerful extended radio sources at their respective redshifts, 
which could account for the peak of maximally spinning holes, or relative dearth 
of sources with lower values of spin mass-energy relative to black hole mass. 
The sample of LINERs have redshifts between about zero and 0.3, and also 
includes sources with maximal 
spin mass-energy relative to black hole mass, and sources with a range of 
values of spin mass-energy parameters. The local sample of AGN (M03) has a  
broad range of black hole spin mass-energy relative
to black hole mass, and does not exhibit a prominent excess of sources with maximal 
spin mass-energy (see section 4), though such a population could become evident with larger 
local samples. Both the LINER and M03 samples include a significant number 
of sources with high values of the spin parameter $F$, which could be due to 
fluctuations in the normalization parameter $g_j$, or fluctuations due to 
non-simultaneous radio and X-ray observations. 
The overall conclusion is that the samples and sources studied provide 
information about those samples, and it is too early to draw conclusions 
regarding the spin mass-energy parameters relative to the black hole mass 
for the full population of supermassive black holes. 
 
\section{Summary \& Conclusions} 
\label{sec:conclusions}

The spin mass-energy of black holes associated with compact and classical 
double radio sources represents a substantial reservoir of energy that can 
be extracted from the black holes, thereby reducing the black hole mass, 
powering the outflow, and impacting the galactic and extra-galactic environment 
of the source. For the sources studied here, the value of $M_{spin}/M_{dyn}$ is
approximately $0.2 \pm 0.1$. Given that all of the sources studied have collimated 
outflows from a black hole system, and that each sample is subject to different 
selection effects, the results presented here provide information about the 
types of sources studied here, and it is too early to tell whether the results  
apply to black holes in general, as discussed in sections 4 and 5. 

The mapping of black hole spin, $j$, to black hole spin mass-energy, $M_{spin}$,  
relative to the total or gravitational black hole mass, $M_{dyn}$, and vice versa 
is non-linear. Most of the sources studied here have 
substantial spin values and hence contribute significantly to the spin 
mass-energy reservoir. The spin energy available, and the energy extracted 
per outflow event provides an upper bound on or estimate of the energy that 
can be input to the galactic and extragalactic environment by spin powered 
outflows, so this is an important quantity to obtain and study. Preliminary 
studies suggest that about a few percent of the spin energy and angular momentum 
momentum of the black hole are extracted per outflow event for FRII AGN (Daly 2011, 2020), 
and as discussed in section 1, it is likely that there are multiple outflow 
events per source.  

The results obtained with the AGN studied here are consistent with 
expectations based on independent studies and numerical simulations, 
and indicate a substantial reservoir of spin mass-energy associated with 
supermassive black holes that power collimated outflows. 
There may also be a population of supermassive black holes 
with very low spin energy relative to black hole mass, since any hole with a 
spin, $j$, less than about 0.5 possesses less than about 10 \% of the maximal 
spin energy  or about 3 \% spin mass-energy relative to black hole mass, and 
thus may not produce an observable powerful outflow. 
The spin of a black hole evolves as a result of mergers, accretion, and 
outflow events, which may well depend upon environment and redshift. 

\section*{Acknowledgments}It is a pleasure to thank the many colleagues with whom 
this work was discussed, especially Jean Brodie,  
Yan-Fei Jiang, Chiara Mingarelli, Masha Okounkova, and 
Rosie Wyse. The Flatiron Institute is supported by the Simons Foundation. 
This work was also supported in part by the Losoncy Fund 
and the PSU Berks Advisory Board,  
and was performed in part at the Aspen Center for Physics, 
which is supported by National Science Foundation grant PHY-1607611.

\end{document}